\documentclass[prb,twocolumn,superscriptaddress,longbibliography]{revtex4-1}
\usepackage{graphicx}
\usepackage{dcolumn}
\usepackage{bm}
\usepackage{amsmath}
\usepackage{amsfonts}
\usepackage{fix-cm}
\usepackage{ulem}

\newcommand{\bra}[1]{\langle #1|}
\newcommand{\ket}[1]{|#1\rangle}

\newcommand{\ex}[1]{\langle #1 \rangle}

\newcommand{\beq}{\begin{eqnarray}}
\newcommand{\eeq}{\end{eqnarray}}

\newcommand{\eqr}[1]{Eq.~(\ref{#1})}
\newcommand{\fig}[1]{Fig.~\ref{#1}}

\newcommand{\be}{\begin{equation}}
\newcommand{\ee}{\end{equation}}
\newcommand{\bq}{\begin{eqnarray}}
\newcommand{\eq}{\end{eqnarray}}

\begin{document}
\title{Amplified and tunable transverse and longitudinal spin-photon coupling in hybrid circuit-QED}
\author{Neill Lambert}
\affiliation{CEMS, RIKEN, Saitama 351-0198, Japan}
\author{Mauro Cirio}
\affiliation{CEMS, RIKEN, Saitama 351-0198, Japan}
\author{Matthieu Delbecq}
\affiliation{CEMS, RIKEN, Saitama 351-0198, Japan}
\author{Giles Allison}
\affiliation{CEMS, RIKEN, Saitama 351-0198, Japan}
\author{Marian Marx}
\affiliation{CEMS, RIKEN, Saitama 351-0198, Japan}
\affiliation{Department of Applied Physics, University of Tokyo, Bunkyo-ku, Tokyo, Japan}

\author{Seigo Tarucha}
\affiliation{CEMS, RIKEN, Saitama 351-0198, Japan}
\affiliation{Department of Applied Physics, University of Tokyo, Bunkyo-ku, Tokyo, Japan}

\author{Franco Nori}
\affiliation{CEMS, RIKEN, Saitama 351-0198, Japan}
\affiliation{Department of Physics, University of Michigan, Ann Arbor, Michigan 48109-1040, USA}

\begin{abstract}
We describe a method to tune, in-situ, between transverse and longitudinal light-matter coupling in a hybrid circuit-QED device composed of an electron spin degree of freedom coupled to a microwave transmission line cavity.  Our approach relies on periodic modulation of the coupling itself, such that in a certain frame the interaction is both amplified and either transverse, or, by modulating at two frequencies, longitudinal. The former realizes an effective simulation of certain aspects of the ultra-strong coupling regime, while the latter allows one to implement a longitudinal readout scheme even when the intrinsic Hamiltonian is transverse, and the individual spin or cavity frequencies cannot be changed. We analyze the fidelity of using such a scheme to measure the state of the electron spin degree of freedom, and argue that the longitudinal readout scheme can operate in regimes where the traditional dispersive approach fails.
\end{abstract}
\maketitle

\section{Introduction}

Electron spin is a highly robust quantum degree of freedom whose use in quantum information is often limited by the difficulty of implementing fast high-fidelity readout and the realization of long-distance interactions\cite{Childress2004,Bergenfeldt2012,Lambert2013,Contreras2013,Bergenfeldt2013}. Spin-photon coupling in hybrid devices composed of double-quantum-dots (DQD) coupled to superconducting transmission-line cavities is being investigated and developed as a means to overcome these difficulties\cite{Delbecq2011,Frey2012,Hu2012,Frey2012b,Petersson2012,Delbecq2013,Bergenfeldt2013,Toida2013,Wallraff2013,Viennot2014,Liu2015,Stockklauser2015,Deng2014,ViennotScience2015,Beaudoin2016}. Very recently several experiments have demonstrated strong spin-photon coupling \cite{Mi2017,Samkharadze2017,Landig2017} based on coupling mediated by the charge degree of freedom \cite{Bruhat2016,Petta2017,Stockklauser2017,Petta2017,Petta2017b}.
In addition to applications in quantum information, such devices harbour new physics, including controllable single-atom lasing~\cite{Savage1992,McKeever2003,Oleg,Ashhab2009}, ground-state lasing~\cite{Cirio2015} bistability~\cite{Lambert2015}, non-equilibrium thermodynamics~\cite{Bergenfeldt2014}, and quantum phase transitions~\cite{Ashhab2013}. 

In this work we focus on the practical task of how to switch, in situ, between an amplified {\it longitudinal~\cite{Didier2015,Royer2016, Zhao2015,Wang2016,Wang2017,Wang2017a}}, and an amplified {\it transverse} coupling, by only modulating the coupling strength, and without changing the spin or cavity energies directly. With the former (amplified longitudinal coupling) one can realize fast high-fidelity readout \cite{Didier2015} and qubit-qubit coupling~\cite{Royer2016}. With the latter (amplified transverse coupling) one can investigate the extreme limits of light-matter coupling \cite{Stassi2017a,Stassi2017} in a simulated manner \cite{Braumuller2016}.

 Our primary result is that one can realize an effective amplified longitudinal coupling even when there is a non-negligible intrinsic transverse term in the Hamiltonian by modulating the coupling strength at {\it both the cavity and qubit frequencies simultaneously} (two-tone), and moving to an appropriate frame. We show that this works optimally when the intrinsic qubit frequency is half of the cavity frequency. The effect can be intuitively understood in terms of  a simultaneous resonant force on the cavity and electron-spin-resonance (ESR) on the qubit. We say that the coupling strength is amplified in the sense that the influence of the qubit on the cavity is increased drastically as the effective cavity frequency is reduced.

 With the electron spin-based devices we discuss in this work this modulation is potentially achievable with electrical control of a single gate-voltage \cite{Jin2012, Royer2016}. This method is particularly desirable when, as is the case we outline below, one cannot (or may not want to) directly engineer  a longitudinal interaction, or cannot control in-situ the intrinsic properties of the device (other than the coupling itself).  The two-tone approach, similar in philosophy to a stroboscopic scheme recently implemented in experiments \cite{Hacohen-Gourgy2016},  also has the advantage that, when used as a means to measure the qubit state, it is faster than dispersive readout, and can still operate well in the limit of strong coupling and a bad cavity. The downside is that, like the normal dispersive readout scheme, it is approximate, and the quantum non-demolition (QND) nature of the measurement breaks down away from ideal parameters (unlike an ideal intrinsic longitudinal coupling). Thus the longitudinal readout part of our proposal lies between the ``pure'' longitudinal case and the traditional dispersive case, with the fast readout of the former, and the potentially easier implementation of the latter (albeit with corresponding limits to its intrinsic QND fidelity away from a sweet spot).

First we describe the basic elements of the spin-photon coupling mechanism.  We then introduce the modulated coupling, and discuss how the two-tone modulation allows us to realize a longitudinal coupling even when the intrinsic Hamiltonian is transverse. We then analyze the fidelity of a two-tone longitudinal measurement scheme, and show how it compares to the normal longitudinal readout (with only a single-tone modulation of the coupling) and dispersive readout approaches.  We finally briefly discuss how a single-tone modulation can give an amplified transverse coupling. In the appendix, we present a detailed analysis of the perturbative limits of the two-tone modulation approach.

\section{Origin of the spin-photon coupling}

Following the approach of Ref.~(\onlinecite{Jin2012}) we consider a model of a DQD operating in the two-electron regime, and at the charge-degeneracy point to minimize dephasing.  A microwave resonator modifies the gate voltage that controls the interdot tunneling, which results in a spin-photon coupling as described below. In addition, the electrons in the dots are subject to an external magnetic field $B_{\mathrm{ex}}=B\hat{z}$, separating the triplet states, $T_{+}=\ket{\uparrow \uparrow}$ and $T_{-}=\ket{\downarrow \downarrow}$, from the triplet state, $T_0=(\ket{\uparrow \downarrow}+\ket{\downarrow \uparrow})/\sqrt{2}$, and the singlet state, $S=(\ket{\uparrow \downarrow}-\ket{\downarrow \uparrow})/\sqrt{2}$. For brevity we neglect reference to the corresponding spatial orbital wavefunctions\cite{Burkard1999,Burkard2000,Pederson2007,Pederson2010} of the electrons in the double dot.  The electrons are also subject to inhomogeneous magnetic fields  $B_{L}$ and $B_{R}$, originating from either inhomogeneous nuclear Overhauser fields or the strong gradient field of a micromagnet.  Here, we define,
\beq
\sigma_z &=& \ket{\downarrow \uparrow}\bra{\downarrow \uparrow}- \ket{\uparrow \downarrow}\bra{\uparrow \downarrow} \equiv \ket{T_0}\bra{S} + \ket{S}\bra{T_0}  \\
\sigma_x &=& \ket{\downarrow \uparrow}\bra{\uparrow \downarrow} + \ket{\uparrow \downarrow}\bra{\downarrow \uparrow}  \equiv \ket{T_0}\bra{T_0} - \ket{S}\bra{S}.
\eeq

Within this restricted two-state subspace the Hamiltonian for the spin states of the dots is given by,
\beq
H_d=\frac{J_0}{2} \sigma_x + \frac{\Delta h}{2} \sigma_z,
\eeq
where $J_0$ is the exchange splitting\cite{Burkard1999,Pederson2007,Pederson2010} and $\Delta h=g_e\mu_B (B_L - B_R)$ the difference in local Zeeman energies.  
In this work we focus on the regime where $\Delta h \gg J_0$.

We assume that the superconducting transmission line is coupled to the interdot tunnel gate.  The vacuum state in the cavity has a non-zero voltage that can modify this barrier, and thus induces a Hamiltonian,
\beq
H_c = \omega_c a^{\dagger}a + J_r \sigma_x (a+a^{\dagger})
\eeq
where $ \omega_c $ is the resonant cavity frequency, and $J_r$ is the spin-photon coupling strength (see below). Reference (\onlinecite{Jin2012})  considers the eigenstates of $H_d$ as the qubit basis, and by applying a large {\it global} magnetic field they propose tuning $J_0\rightarrow 0$, to maximize the {\it transverse} spin-boson coupling.  One downside to this approach, however, is that this mechanism of tuning $J_0$ to zero is, to our knowledge, as yet unobserved in experiments. It also requires strong external magnetic fields, which, depending on design, may be incompatible with the critical field-requirements of a superconducting transmission line resonator, and may also reduce the intrinsic strength of $J_r$. In addition, reference (\onlinecite{Jin2012}) suggests that the opposite longitudinal regime can be reached by tuning the Zeeman splitting $\Delta h$, such that it is much smaller than the exchange splitting $J_0$.

Here, we investigate a complimentary approach to this notion of switching between longitudinal and transverse interactions, based purely on modulation of the coupling strength between the cavity and double quantum dot. As mentioned in the introduction, this also allows us to realize fast longitudinal-coupling readout \cite{Didier2015}. This on-chip tunability is particularly beneficial to certain double quantum dot devices where it may be difficult to tune $\Delta h$ in situ., and where an inherently large $\Delta h$ may be desirable for state-preparation purposes.

\subsection{Driven Coupling}

In Ref.~(\onlinecite{Jin2012}) the functional dependence of the exchange-splitting mediated spin-photon coupling $J_r$ is given by:
\beq
J_r(t) = eV_r \sinh \left[ \frac{16 V_h(t) (\omega_0^2 + 2 \omega_L^2)}{\hbar \omega_0^2 \sqrt{\omega_0^2 + \omega_L^2}}\right]^{-1}
\eeq
where $\omega_L = eB /2m$ is the Larmor frequency, $\omega_0$ is the frequency of the harmonic well defining each dot, and $V_h$ is the height of the tunnel barrier between the two dots.  Essentially, the vacuum-fluctuation induced voltage $V_r$ modifies the height of the tunnel barrier, which in turn changes the exchange splitting between triplet and singlet states \cite{Burkard1999,Pederson2007,Pederson2010}.  The height $V_h$ is in practice a tunable parameter which can be controlled by a gate voltage.  By applying time-dependent driving \cite{Medford2013} to this gate voltage, $V_h(t)$, one can make $J_r(t)$ time dependent.  One caveat is, in the same stroke, we also induce a time-dependence in the exchange splitting, $J_0$, itself. However, as discussed in the different context of superconducting qubits \cite{Didier2015}, this type of imperfection has a minimal influence of the fidelity or QND-ness of the measurement.

Using exchange splitting to realize modulated coupling is not the only potential way to implement this tunable spin-photon coupling scheme. Following the proposal described in Ref.~(\onlinecite{Hu2012}) one could couple the spin of a single electron in a double dot structure to the microwave cavity by applying a strong magnetic field gradient with a micromagnet \cite{Mi2017,Samkharadze2017}. This could then be made time dependent by electrical control of the dot potential \cite{Pioro-Ladriere2008}, or modulation of the field gradient with a suspended nano-magnet \cite{Lambert2008a}.    There are various advantages and disadvantages to using single spin versus a effective singlet-triplet qubit. The latter tends to have worse dephasing than the former when the exchange splitting or the dot bias are changed \cite{Hu2012}, but has the advantage of being well developed in terms of electrical preparation and readout of the qubit state.

\begin{figure}[!t]

\includegraphics[width=1.0\columnwidth]{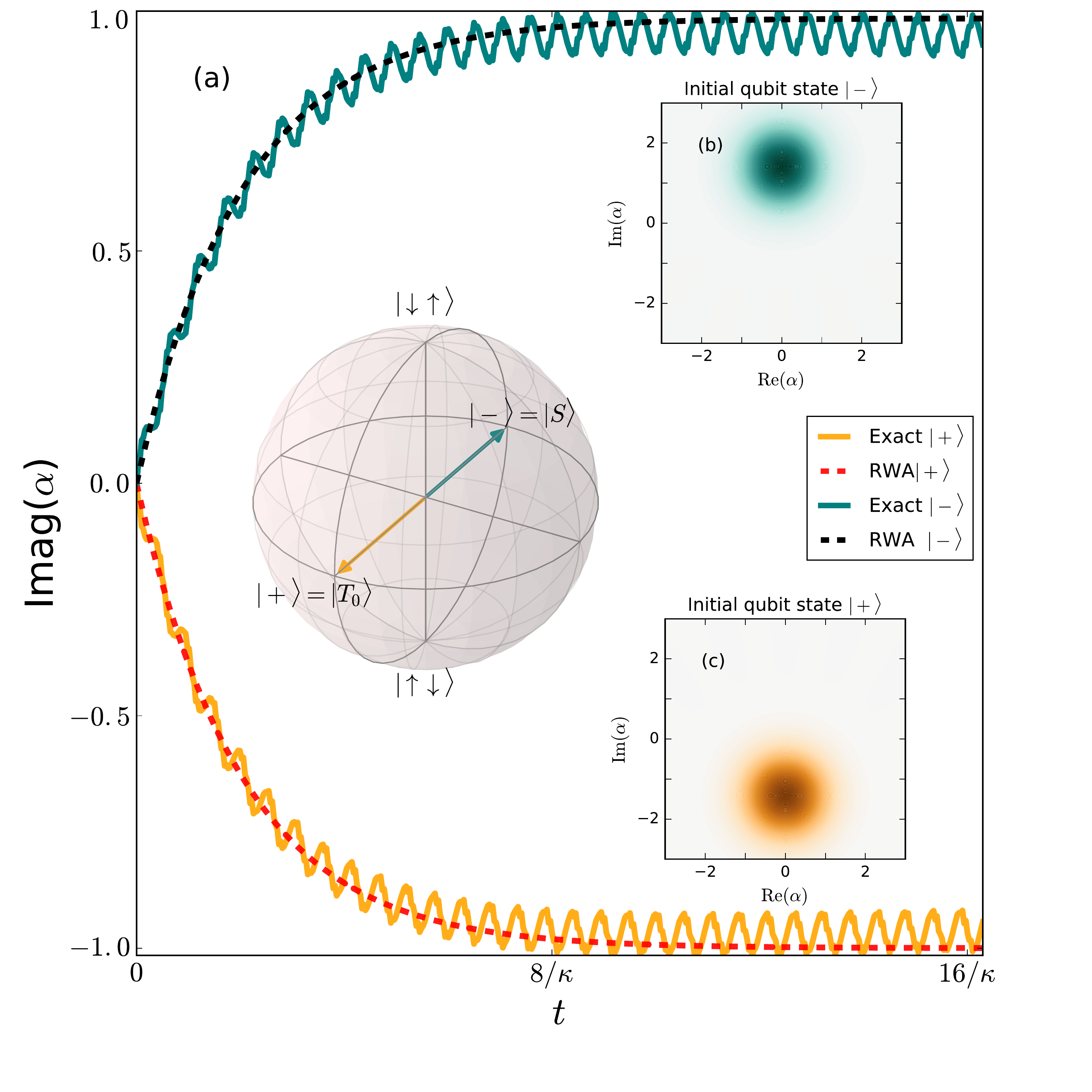}
\caption{(a) Shows an example of the evolution of the imaginary part of the cavity state $\alpha =\ex{a}$ when the qubit is prepared in the $\ex{+}$ or $\ex{-}$ eigenstates of the $\sigma_x$ for the two-tone modulated longitudinal readout scheme. Note that in our notation, the scheme is longitudinal in the $\sigma_x$ basis, and is quasi-QND in that basis, as shown by the Bloch sphere inset. The dashed curves show the approximate RWA solution, while the solid lines show the full numerics, which includes oscillations due to the counter-rotating terms.  The insets (b) and (c) show the Wigner function of the cavity state for the different initial qubit states at $t=16/\kappa$.  In this figure we have used non-ideal parameters to accentuate the unwanted oscillations, with  $\Delta h=0.15 \omega_c$,  $J_r=0.05{\omega_c}$, and $\kappa= J_r/2$. The oscillations due to counter-terms reduce the QND fidelity of the measurement, but this can be improved by of course increasing $\Delta h$ or reducing the coupling strength $J_r$ [and correspondingly reducing $\kappa$ to maintain the same signal magnitude $\alpha(t\rightarrow \infty)=J_r/2\kappa$.] }
 \label{fig1}
\end{figure}

\section{Two-tone driving and amplified longitudinal readout}

When $\Delta h \gg J_0$ our intrinsic Hamiltonian is transverse, and we assume $\Delta h$ is a static property that cannot be tuned in-situ.  However, as mentioned in the introduction, we can access an effective amplified longitudinal regime by driving the coupling at two frequencies.  When the natural splitting of the qubit and cavity are off-resonance ($\Delta h \sim \omega_c/2$), we can do quasi-QND amplified longitudinal readout of the $\sigma_x$ basis, as we have defined it. More specifically,  returning again to the Hamiltonian
\beq
H= \omega_c a^{\dagger}a +\frac{\Delta h}{2} \sigma_z+ J_r(t) \sigma_x (a+a^{\dagger})   \label{H_QND4}
\eeq
and choosing
\beq J_r(t)=J_r\cos \left(\omega_c t\right)\cos \left(\Delta h t\right)\eeq
and moving to a rotating frame under the unitary transformation $U = \exp{i (\omega_c a^{\dagger}a   + (\Delta_h/2) \sigma_z) t}$, under the assumptions that $\omega_c, \Delta h \gg J_r$, and neglecting fast oscillating terms~\cite{Ashhab2007} of frequency $2 \omega_c$, $2 \Delta h$, $\omega_c +\Delta h$, and $\omega_c - \Delta h$, we obtain
\beq
H_{0}=\frac{J_r}{4}\sigma_x (a+a^{\dagger})\;.
\label{H_QND5}
\eeq
Thus, we have effectively entered a frame where both the cavity frequency (as in the previous section) and the qubit splitting are zero.  This Hamiltonian thus describes a $\sigma_x$-dependent resonant force on the cavity, and with it we can perform fast quasi-non-QND readout of the eigenstates of that basis (albeit in a rotating frame).  It is more traditional to redefine the basis states to measure in $\sigma_z$, but we refrain from doing so.  With this  Hamiltonian the qubit-dependent displacement of the cavity tends towards \beq \alpha=\ex{a}={\pm}J_r/2 i\kappa\eeq in the steady-state, as shown in \fig{fig1}, and does so faster than the equivalent dispersive interaction \cite{Didier2015} (here $\kappa$ is the cavity loss rate, see below for a full description).
If one prefers to perform a measurement in the $\sigma_z$ basis one must of course initially apply a rotation on the qubit before the measurement is performed.


The regime of validity of \eqr{H_QND5} depends strongly on $\omega_c, \Delta h \gg J_r$ and $\omega_c - \Delta h \gg J_r$, implying an optimal point of $\omega_c = 2 \Delta h$. This is easily analyzed using Van-Vleck perturbation theory, as shown in the appendix.  In particular, in the regime $\frac{J_r}{8} < \Delta h < \omega_c - \frac{J_r}{8}$ the lowest order non-QND terms arising from the perturbation theory are:
\beq
\label{eq:VVmain}
H^{\text{VV}}&=& H_{0}+\left(\frac{J_r}{4}\right)^2\left[\frac{(a+a^\dagger)^2}{2\Delta h}\right.\\
&-&\left.\frac{\Delta h}{{\omega_c}^2-\Delta h^2}\left(a^\dagger a+\frac{1}{2}\right)\right]\sigma_z,\nonumber
\eeq

We validate this analysis with a numerical simulation of the full dynamics \cite{Johansson2012,Johansson2013}, which involves solving a Master equation including the full time-dependent Hamiltonian \eqr{H_QND4} and cavity loss rate $\kappa$,
\beq
\dot{\rho}= -\frac{i}{\hbar}[H(t),\rho] + \frac{\kappa}{2}\left[2a\rho a^{\dagger} - a^{\dagger}a\rho -   \rho a^{\dagger}a \right].
\eeq
Here we neglect qubit (DQD) loss and dephasing, and focus only on the influence of the cavity losses.

\begin{figure}[t]

\includegraphics[width=1.0\columnwidth]{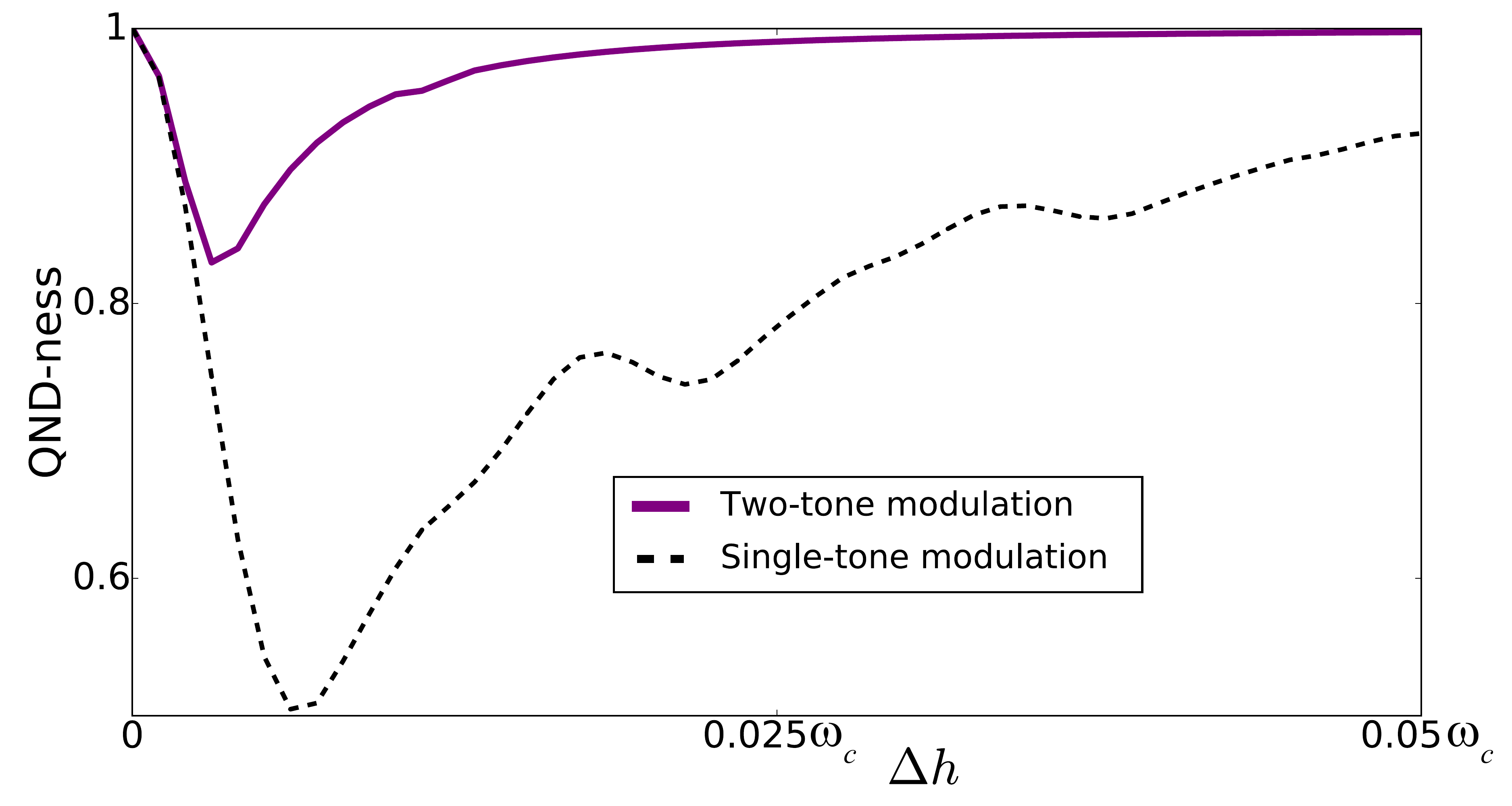}
\caption{As a figure of merit of the QND fidelity of the measurement process we use, for an initial excited state $\vert + \rangle$, i.e., Min$[\ex{\vert + \rangle \langle + \vert}]_{\tau}$, and we take the maximum evolution time as $\tau=2/\kappa$.  Here we use parameters closer to those expected in a DQD-Cavity setup, with $\omega_c = 5$ GHz, $J_r=50$ MHz and $\kappa=25$ MHz. We tune $\Delta h$ across the range $0$ to $250$ MHz. When $\Delta h=0$ we recover the pure longitudinal result of Didier \textit{et al.} \cite{Didier2015}.  As $\Delta h$ is increased the QND-ness of the two-tone modulation scheme decreases until a critical turning point, corresponding a passage from an adiabatic regime to a fast modulation regime, where a RWA starts to become valid. This regime is ideal when $\Delta h= \omega_c/2$ (see appendix for a complete analysis), but we see that,  for the parameters in this example, it already performs well as $\Delta h \rightarrow \omega_c/20$.}
\label{fig2}
\end{figure}

 Figures of merit for the efficiency of the readout scheme are the non-destructiveness (QND-ness) and the time-dependent signal-to-noise ratio.
In \fig{fig2}, from the full numerical results, we show a simple measure,  Min$[\ex{\vert + \rangle \langle + \vert}]_{\tau}$, of the non-destructiveness of the measurement in terms of the minimum overlap between the state of the qubit (in the rotating frame) and the initial state $\vert + \rangle$,  across the whole time evolution interval $\tau$, as a function of $\Delta h$. At $\Delta h=0$, we retrieve the purely longitudinal results of Didier \textit{et al.} \cite{Didier2015}.  As $\Delta h$ increases, readout relying on a single-tone modulation of the coupling just at the cavity frequency of course fails to produce a satisfactory QND-ness. However, by modulating at two frequencies (solid curve) we observe first a drop in the QND-ness, and then right afterwards we see a revival, as the simplified RWA model \eqr{H_QND5}, which predicts ideal non-destructive measurement at $\Delta h =\omega_c/2$, becomes valid (see appendix).

The time-dependent signal-to-noise ratio is given by,
\beq
\mathrm{SNR}(\tau) = \frac{ \ex{M(\tau)_{+}} - \ex{M(\tau)_{-}}}{\left[\ex{\delta M(\tau)_{+}^2}+\ex{\delta M(\tau)_{-}^2}\right]^{1/2}},\label{defSNR}
\eeq
where ${+}$ or ${-}$ refers to  the qubit-state in the $\sigma_x$ basis, and
\beq
M(\tau) = \sqrt{\kappa}\int_0^{\tau} dt [a_{\mathrm{out}}^{\dagger}(t)+a_{\mathrm{out}(t)}],
 \eeq is the homodyne signal in terms of the integrated  quadrature amplitude of photons leaking out of the cavity at a rate $\kappa$ (where $a_{\mathrm{out}}(t) = \sqrt{\kappa} a(t) +  a_{\mathrm{in}}(t) $ includes vacuum noise $\ex{a_{\mathrm{in}}(t)a_{\mathrm{in}}^{\dagger}(t')} = \delta(t-t')$). The integrated noise is given by the sum of the variance of both outcomes, $\delta\! M(\tau)=M(\tau) - \ex{M(\tau)}$, which can be evaluated as~\cite{Wiseman2010,Fan2013},
\beq
\delta\! M(\tau)^2&=&\kappa^2\!\!\int_0^{\tau} \!\! \!\!dt \int_0^{\tau} \!\! \!\!dt' \left(\mathrm{Tr}[(a+a^{\dagger})\exp\left\{\mathcal{L} (t'-t)\right\}\right. \nonumber\\
&& (a\rho(t) + \rho(t)a^\dagger) ]u(t'-t)\\
&&+\mathrm{Tr}[(a+a^{\dagger})\exp\left\{\mathcal{L} (t-t')\right\}\nonumber\\
&&\left. (a\rho(t') + \rho(t')a^\dagger) ]u(t-t') \right)
+ \kappa \tau - \ex{M(\tau)}^2\nonumber
\eeq
which, in the case that the state in the cavity is a coherent state, reduces to $\delta\! M({\tau})^2=\kappa {\tau}$,  where $\tau$ is the total measurement period.

\begin{figure}[t]

\includegraphics[width=1.0\columnwidth]{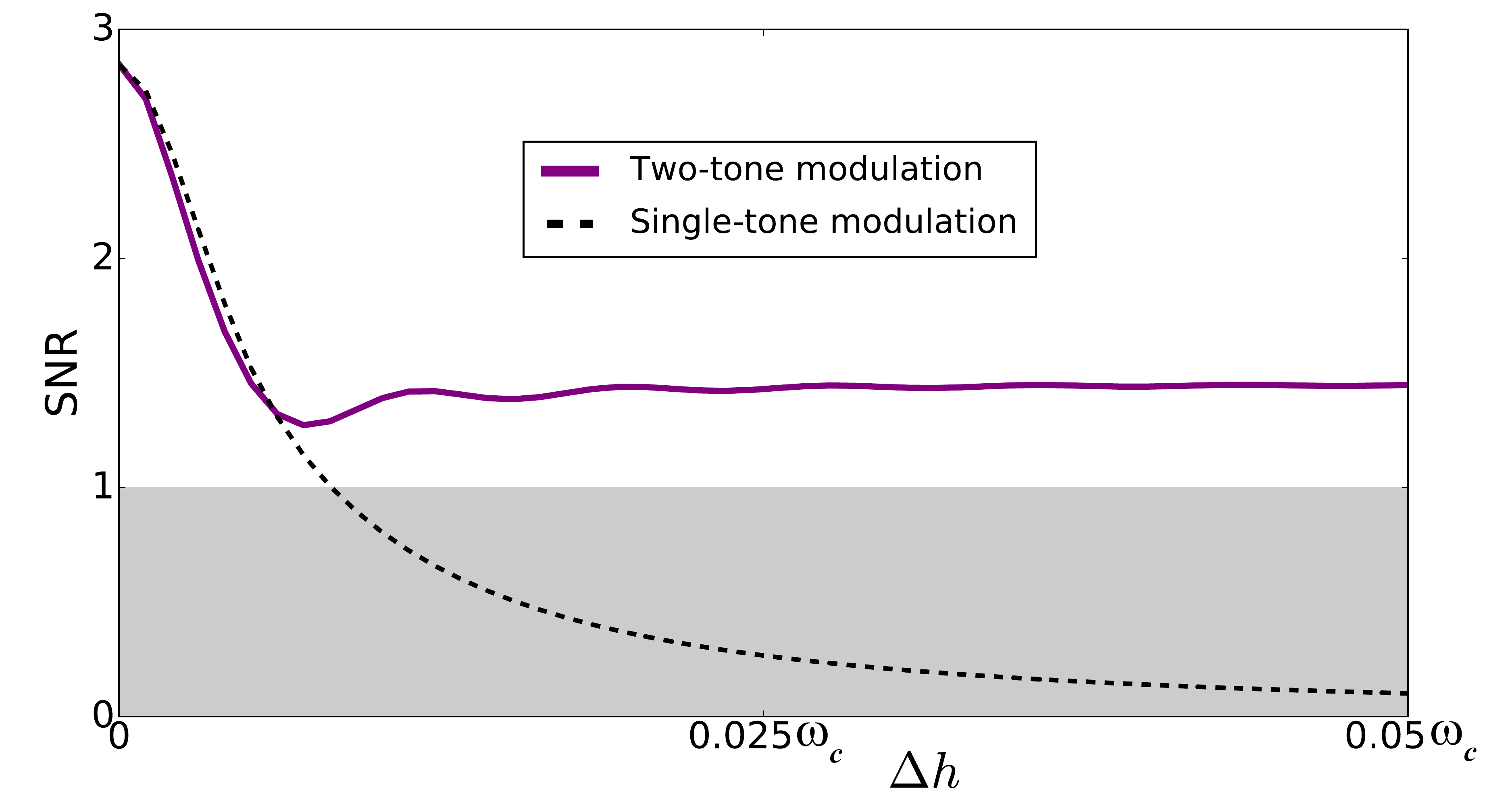}
\caption{Here we show integrated signal-to-noise ratio obtained  up to time  $\tau=2/\kappa$, as a function of $\Delta h$, with other parameters set as in \fig{fig2}. The SNR is maximal for $\Delta h\rightarrow 0$, then drops and saturates as $\Delta h$ is increased.}
\label{fig3}
\end{figure}

In \fig{fig3} we show the signal-to-noise ratio, \eqr{defSNR},  also as a function of $\Delta h$, up to a  maximum integration time of ${\tau}=2/\kappa$. The larger SNR at $\Delta h=0$ is ultimately due to the effectively larger coupling $J_r$, compared to the case when one has a finite frequency of modulation (i.e., at $\Delta h=0$, $J_r$ is effectively two times larger compared to when the modulation at finite $\Delta h$ occurs, and whence averaging over fast oscillations effectively reduces the coupling strength). As $\Delta h$ increases, as with \fig{fig2}, modulating the coupling at just a {\it single} frequency is accompanied with a loss of signal.  However, if one modulates at {\it two} frequencies, $\omega_c$ and $\Delta h$, the SNR plateaus, as expected from \eqr{H_QND5} and the analysis performed in reference (\onlinecite{Didier2015}).

In comparing their pure longitudinal measurement scheme  to the traditional dispersive approach, reference (\onlinecite{Didier2015}) argued that the SNR of the longitudinal scheme increases faster than that of the dispersive one at short times: \beq
\text{SNR}(\tau) &\propto & \frac{1}{\kappa}(\kappa \tau)^{5/2}\quad \text{for dispersive case},\nonumber\\
\text{SNR}(\tau) &\propto & \frac{1}{\kappa}(\kappa \tau)^{3/2}\quad \text{for longitudinal readout},\nonumber\\
 \text{SNR}(\tau) &\propto & \frac{1}{\kappa}(\kappa \tau)^{1/2}\quad \text{for both at longer times}\quad \tau \gg \kappa^{-1}.\nonumber
 \eeq  While this is also the case for the two-tone readout, we point out an additional advantage of both the purely longitudinal scheme  \cite{Didier2015}  and that of our two-tone modulation readout. In the examples shown in Figs. \ref{fig2} and \ref{fig3} we evolve to time scales of order $\kappa^{-1}$ and we set the loss $\kappa = J_r$. The choice of this ratio is important in the sense that a smaller coupling would give a lower magnitude steady state, and a smaller SNR, while a smaller loss $\kappa$ would give a slower overall readout time.  In the normal dispersive readout, the equivalent requirement for a non-negligible SNR on this same time scale is
\beq
\frac{E}{\kappa}~\frac{J_r^2}{\Delta} > \kappa/2,
 \eeq where $\Delta = \omega_c - \Delta h$, and $E$ is the magnitude of an external resonant drive on the cavity.  However, due to the perturbative nature of the dispersive interaction, there is a limit on the value of $E/\kappa < (\Delta/\sqrt{8} J_r)$ (sometimes termed the ``critical photon number'' \cite{You2003,blais04,You2005,blais08,blais09,You2011}). This in turn limits the value of $\kappa$ one can allow in the dispersive readout scheme at least to $J_r/\sqrt{2}$, and in practice much less (the critical photon number is an extreme upper limit, related to how dressed the eigenstates of the dispersive Hamiltonian become at larger photon numbers). On the other hand, the longitudinal schemes function with high fidelity up to the ``bad cavity'' limit of $\kappa = J_r$, a regime which potentially offers faster readout.  For example, for the same parameters we use in the figures, the dispersive readout fails completely. 

\section{Amplified transverse coupling regime}

The magnitudes of the spin-photon coupling strengths predicted in theory \cite{Hu2012, Jin2012}, and seen in experiments so far \cite{Mi2017,Samkharadze2017,Landig2017}, are in the strong coupling regime (in that it exceeds the qubit and cavity losses).  However, they are still far from the ultra-strong regime \cite{Ashhab2010, Stassi2017a,Stassi2017}, as they are orders of magnitude smaller than the qubit or cavity frequency themselves.
In addition, in the system we describe in this paper, the singlet-triplet spin-qubit is typically off-resonant with the cavity. If one wishes to realize effective resonant interactions, or even simulate \cite{Ballester2012,Grimsmo2013,Braumuller2016} certain aspects of the ultra-strong coupling regime, one can do so by modulating the qubit-cavity coupling, $J_r(t)$ to make the influence of the qubit on the cavity again akin to a resonant force.  One can do this  by now choosing \beq J_r(t) = J_r \cos\left(\omega_d t\right)\eeq
In which case, the total Hamiltonian becomes
\beq
H&=&\frac{J_0}{2} \sigma_x + \frac{\Delta h}{2} \sigma_z + \omega_c a^{\dagger}a + J_r \cos(\omega_d t) \sigma_x (a+a^{\dagger}).\nonumber
\eeq
Applying a standard transformation $U = \exp i \omega_d a^{\dagger}a  t$, this Hamiltonian becomes
\beq
H&=&\frac{J_0}{2} \sigma_x + \frac{\Delta h}{2} \sigma_z + (\omega_c-\omega_d) a^{\dagger}a \\
&+& J_r \cos(\omega_d t) \sigma_x (ae^{-i  \omega_d t}+a^{\dagger}e^{i  \omega_d t}).\nonumber
\eeq
Applying the rotating wave approximation (RWA), assuming  $\Delta h, J_r \ll \omega_d$, in the limit that $J_0$ is negligible, one obtains,
\beq
H^{R}= \frac{\Delta h}{2} \sigma_z + (\omega_c-\omega_d) a^{\dagger}a + \frac{J_r}{2}  \sigma_x (a+a^{\dagger}).
\eeq
For resonant interactions, one can choose $(\omega_c-\omega_d) = \Delta h$.   As the effective cavity frequency is reduced, the influence of the qubit on the cavity is amplified. To realize certain aspects of the ultra-strong coupling regime one can choose $(\omega_d - \omega_c)=0$, thus, as in the longitudinal case, entering a frame where the cavity frequency vanishes.
 In principle, this would also allow one to study a non-equilibrium variant of the single-qubit Dicke phase transition \cite{Ashhab2013},  similar to the non-equilbrium Dicke phase transition model studied by Bastidas \textit{et al.} \cite{Bastidas2012}.

\section{Conclusions}

In this work we showed how a two-tone modulation of the coupling between a qubit, as exemplified with the singlet-triplet states in a double quantum dot, and a cavity allows one to switch between transverse and longitudinal coupling schemes. While being more ``approximate" than a purely engineered longitudinal coupling, and thus not perfectly QND in some regimes, this approach allows one to switch between transverse and  longitudinal coupling, as required.  For the latter, we presented a detailed perturbative analysis in the Appendix, to show the robustness of the scheme for realistic parameters. Finally, we argued that the longitudinal scheme can be used in the ``bad cavity'' (large $\kappa$) limit, in principle allowing for a faster readout.  Of course, this approach can also be applied to traditional circuit QED \cite{Didier2015}, and perhaps also to other approaches to spin-photon coupling \cite{Hu2012, Mi2017,Samkharadze2017}.

\acknowledgements

We acknowledge discussions with Juan Rojas-Arias.  FN was partially supported by the MURI Center for Dynamic Magneto-Optics via the AFOSR Award No. FA9550-14-1-0040, the Japan Society for the Promotion of Science (KAKENHI), the IMPACT program of JST, JSPS-RFBR grant No 17-52-50023, CREST grant No. JPMJCR1676.  NL and FN acknowledge support from RIKEN-AIST Challenge Research Fund, and the Sir John Templeton Foundation.

\section{Appendix}

\begin{figure}[!t]

\includegraphics[width=1.0\columnwidth]{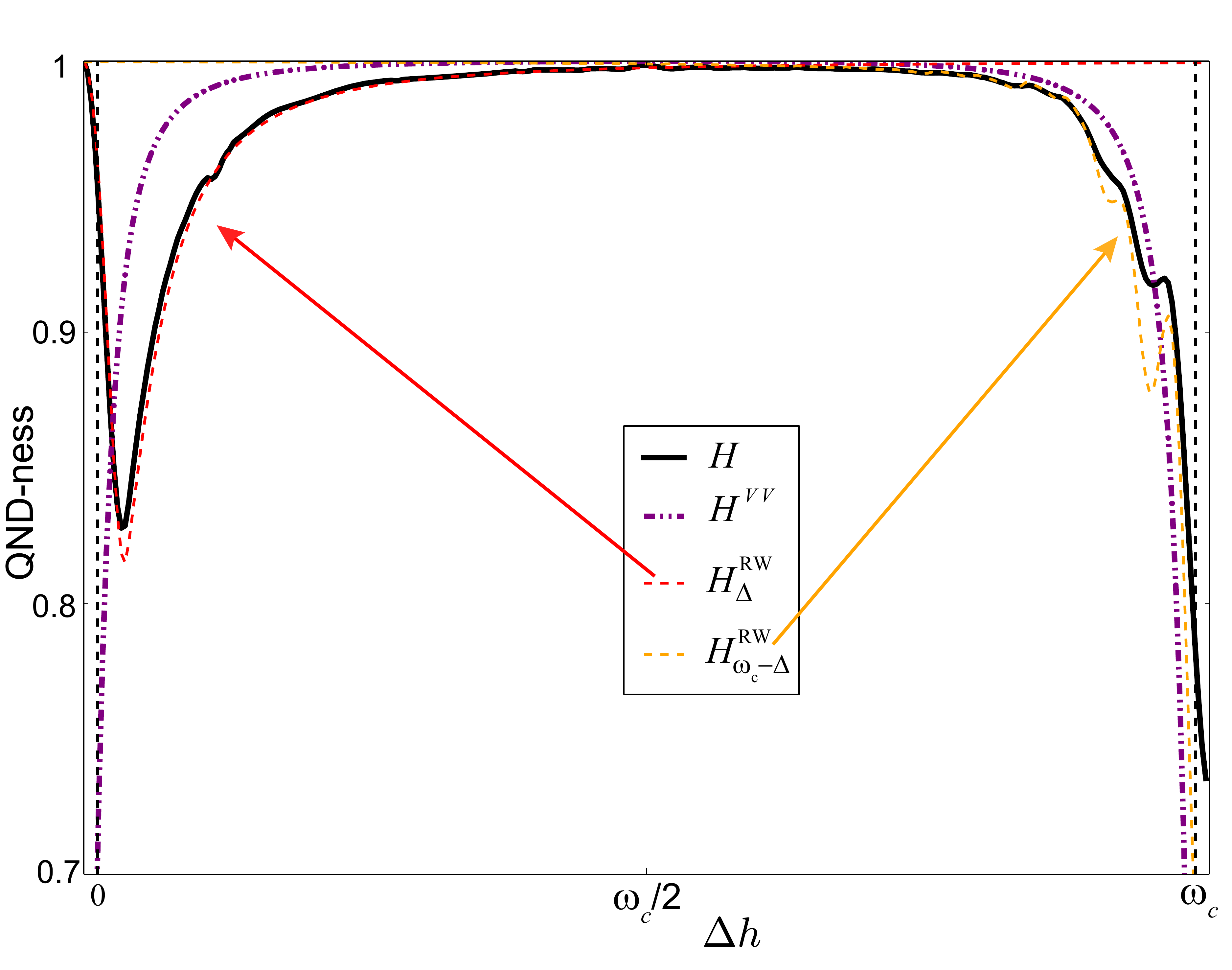}
\caption{As in figure \ref{fig2} we plot the QND fidelity of the measurement process QND fidelity Min$[\ex{\vert + \rangle \langle + \vert}]_{\tau}$ for an initial excited state $\vert + \rangle$ as a function of $\Delta h$.  In this figure we use larger coupling magnitude, $J_r=0.1{\omega_c}$, and loss $\kappa =  J_r/2$, to accentuate the deviation from the ideal QND behavior. As before we choose the total integration time $\tau=2/\kappa$. In black is the result for the full system Hamiltonian while the other lines correspond to the effective Hamiltonians listed in the table in the appendix. The black dashed vertical lines represent the points $\Delta h=\tilde{J}_r, {\omega_c}-\tilde{J}_r$ which set the boundary between the adiabatic and high-frequency regimes (which, for the reasons explained in the text, we expect to be valid in the long measurement time limit, i.e., as $\kappa \tau\rightarrow\infty$). It is clear that operating at the point $\Delta h = \omega_c/2$ is optimal, as discussed in our perturbative analysis, apart from the point corresponding to $\Delta h=0$, where the Hamiltonian is intrinsically longitudinal (corresponding to the proposal in reference (\onlinecite{Didier2015})).}
\label{fig4}
\end{figure}

\begin{figure}[!t]

\includegraphics[width=1.0\columnwidth]{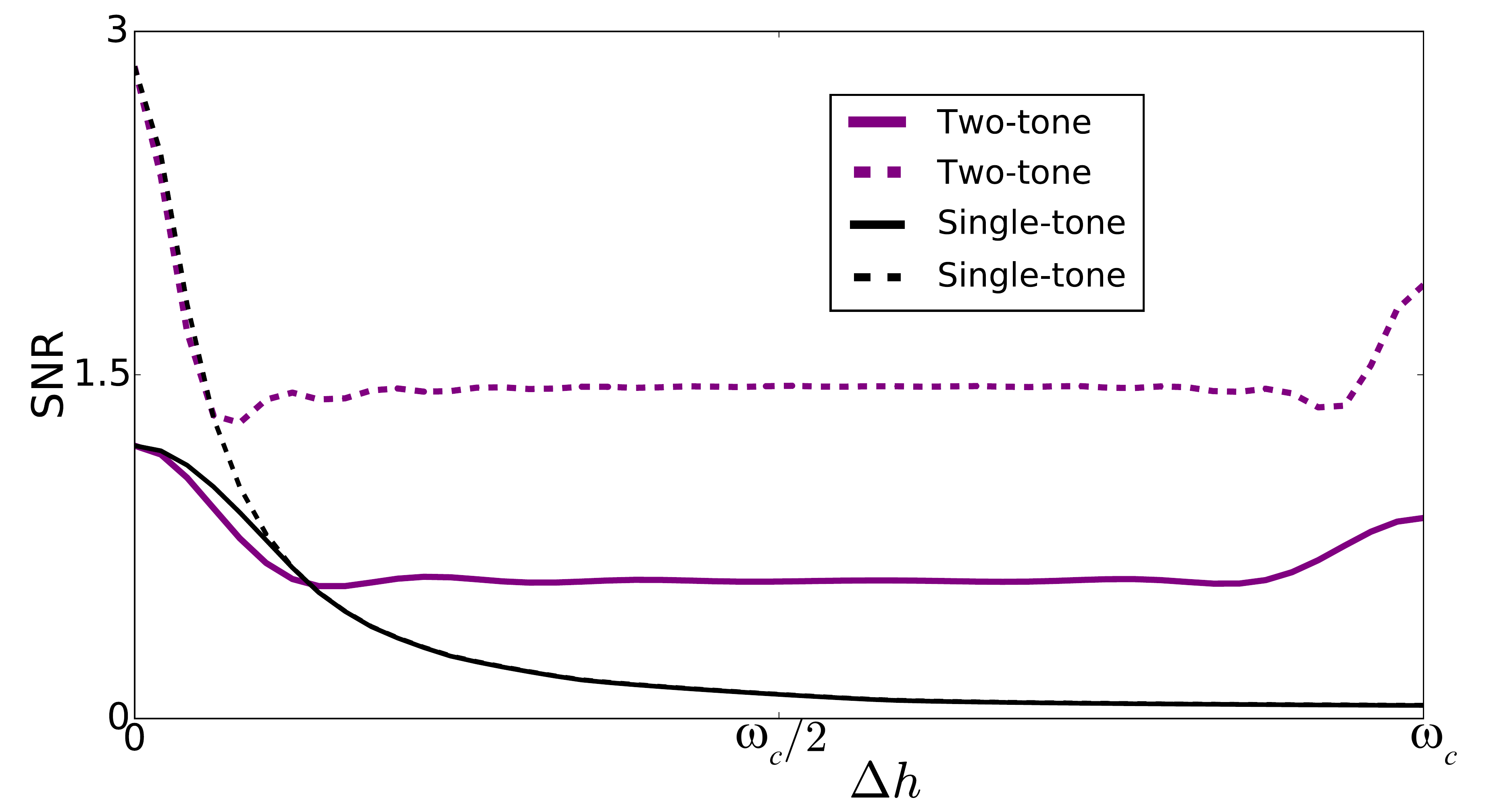}
\caption{For completeness, as in figure \ref{fig3} we show the SNR as a function of $\Delta h$, with other parameters as in \fig{fig4}. As expected, the SNR is large at $\Delta h = \omega_c/2$, and maximal for $\Delta h\rightarrow 0$. The solid lines are for $\tau = 1/\kappa$ while the dashed lines are for $\tau=2/\kappa$, illustrating how one acquires more signal for longer measurement periods.}
\label{fig5}
\end{figure}

In this appendix we present a perturbative analysis which explains the different features of \fig{fig2} and \fig{fig3}.  It is helpful to first show the full behavior of the QND-ness, \fig{fig4}, and SNR, \fig{fig5}, for a full range of $\Delta h$ from $0$ to $\omega_c$. In these figures we user large coupling and loss rates, to increase the error in the RWA approximation, and make it more visible for comparison to the analytical analysis in the next section.

\subsection{Perturbative analysis}

Stating from the full Hamiltonian,
\begin{equation}
\begin{array}{lll}
H=\frac{\Delta h}{2}\sigma_z+\omega_c a^\dagger a + J_r \cos{(\omega_c t)}\cos{(\Delta h t)}\sigma_x (a+a^\dagger),
\end{array}
\end{equation}
with two-tone modulation of the coupling, we can perform a perturbative analysis of the different regimes as we tune $\Delta h$.

It is convenient to write the previous Hamiltonian in a frame $\tilde{\ket{\Psi}}=U\ket{\Psi}$, with $U=\exp{[i(\omega_c a^\dagger a+ (\Delta h/2) \sigma_z )t]}$ as
\begin{equation}
\renewcommand*{\arraystretch}{2}\begin{array}{lll}
H &=& J_r \cos{(\omega_c t)}\cos{(\Delta h t)}(e^{i\Delta h t}\sigma_++e^{-i\Delta h t}\sigma_-)\\
&\times&(e^{i\omega_c t}a^\dagger+e^{-i\omega_c t}a)\\
&=&H_0+\displaystyle\sum_{n_{\Delta h},n_C=-1,0,1}H_{nk}~e^{2i(n_{\Delta h}\Delta h+n_C{\omega_c})}
\end{array}
\end{equation}
where
\beq
H_{0}&=&\frac{g}{4}\sigma_x (a+a^\dagger),\\
H_{1,0}&=&\frac{J_r}{4}(a+a^\dagger)\sigma_+,\nonumber\\
H_{0,1}&=&\frac{J_r}{4}\sigma_x a^\dagger,\nonumber\\
H_{1,1}&=&\frac{J_r}{4}\sigma_+a^\dagger, \quad \mathrm{and} \nonumber\\
H_{-1,1}&=&\frac{J_r}{4}\sigma_- a^\dagger\nonumber
\eeq
with $H_{-n,-k}=H^\dagger_{n,k}$, and $H_{0,0}=0$.\\

In the following we assume ${J_r}/{\omega_c}<1$, and, for formal convenience, define
\begin{equation}
\tilde{J}_r=\frac{J_r}{8}
\end{equation}
Then, the Hamiltonian can be simplified by performing either
\begin{itemize}
\item A rotating wave approximation (RWA) which allows us to neglect terms which rotate at a frequency $\omega$ satisfying $\lambda_{\text{RWA}}(\omega)={\omega}/{\tilde{J_r}}<1$.
\item An adiabatic approximation (A) which allows to neglect slowly-rotating terms at frequency $\omega$ satisfying $\lambda_{\text{A}}(\omega)={\tilde{J_r}}/{\omega}={1}/{\lambda_{\text{RWA}}}<1$.
\end{itemize}
Specifically, we can analyze the regimes $0<\Delta h<{\omega_c}/2$ and ${\omega_c}/2<\Delta h<{\omega_c}$ separately.

\subsubsection{Regime $0<\Delta h<{\omega_c}/2$}
When $0<\Delta h<{\omega_c}/2$, the RWA allows us to write
\begin{equation}
H^\text{RWA}=H_0+2\tilde{J}_r(a+a^\dagger)(e^{2i\Delta h t}\sigma_++e^{-2i\Delta h t}\sigma_-)
\end{equation}
To proceed further, we need to analyze the perturbative parameters $\lambda_\text{RWA}$ and $\lambda_A$ for the time dependent part of the previous Hamiltonian.
\begin{itemize}
\item When $0<\Delta h<\tilde{J_r}$, we have $\lambda_{\text{RWA}}={\Delta h}/{\tilde{J_r}}>1$ and $\lambda_A={1}/{\lambda_\text{RWA}}<1$, which allows us to perform an adiabatic approximation to obtain
\begin{equation}
H_\text{eff}=H_0 + O(\Delta h)
\end{equation}
The quality of this approximation degrades as $\Delta h\rightarrow\tilde{J_r}$.
\item When $\Delta h=\tilde{J_r}$, the frequency of the time-dependent term becomes equal to its energy scale and $\lambda_\text{RWA}(\tilde{J_r})=\lambda_A(\tilde{J_r})$ and neither a further RWA or the adiabatic approximation are allowed.
\item When $\tilde{J_r}<\Delta h<{\omega_c}/2$ we have $\lambda_{\text{RWA}}={\Delta h}/{\tilde{J_r}}<1$ and $\lambda_A={1}/{\lambda_\text{RWA}}>1$, which allows us to perform a further RWA to get
\begin{equation}
H_\text{eff}=H_0 + O\left(\frac{\tilde{J_r}^2}{\Delta h}\right)
\end{equation}
The corrections in the above will be analyzed below.
\end{itemize}

\subsubsection{ Regime ${\omega_c}/2<\Delta h<{\omega_c}$}
When ${\omega_c}/2<\Delta h<{\omega_c}$, the RWA allows us to write
\begin{equation}
H^\text{RWA}=H_0+2\tilde{J}_r(a+a^\dagger)(e^{2i({\omega_c}-\Delta h) t}\sigma_++e^{-2i({\omega_c}-\Delta h) t}\sigma_-)
\label{RWA1}
\end{equation}
Again, to proceed further, we need to analyze the perturbative parameters $\lambda_\text{RWA}$ and $\lambda_A$.
\begin{itemize}
\item When ${\omega_c}/2<\Delta h<{\omega_c}-\tilde{J_r}$ we have $\lambda_{\text{RWA}}={\Delta h}/{\tilde{J_r}}<1$ and $\lambda_A={1}/{\lambda_\text{RWA}}>1$, which again allows us to perform a further RWA to get
\begin{equation}
H_\text{eff}=H_0 + O\left(\frac{\tilde{J_r}^2}{\Delta h}\right)
\end{equation}
\item When $\Delta h={\omega_c}-\tilde{J_r}$ the frequency of the time-dependent term becomes equal to its energy scale and $\lambda_\text{RWA}({\omega_c}-\tilde{g})=\lambda_A({\omega_c}-\tilde{J_r})$ and once again neither a further RWA or adiabatic approximation are allowed.
\item When ${\omega_c}-\Delta h<\Delta h<{\omega_c}$  we have $\lambda_{\text{RWA}}={({\omega_c}-\Delta h)}/{\tilde{J_r}}>1$ and $\lambda_A={1}/{\lambda_\text{RWA}}<1$, which once again allows us to perform an adiabatic approximation to get
\begin{equation}
H_\text{eff}=H_0 + O({\omega_c}-\Delta h)
\end{equation}
The quality of this approximation degrades as $\Delta h\rightarrow{\omega_c}-\tilde{J_r}$.
\end{itemize}

\subsection{High-Frequency Regime}
Deep in the high-frequency regime, where the condition $\tilde{J_r}\ll\Delta h\ll{\omega_c}-\tilde{J_r}$ is satisfied, all time dependent contributions to the original Hamiltonian $H$ satisfy $\lambda_\text{RWA}\ll1$ and a more rigorous analysis can be performed. By using Van Vleck perturbation theory in Floquet space \cite{Leskes2010,RevModPhys.89.011004} an alternative effective Hamiltonian (see Eq. (\ref{eq:VVmain}) in the main text) can be written as
\begin{widetext}
\begin{equation}
\label{eq:VV}
\renewcommand*{\arraystretch}{2.2}\begin{array}{lll}
H^\text{VV}&=& DH D^{-1}\\
&=&H_0-\frac{1}{2}\displaystyle\sum_{n_{\Delta h} n_C=-1,0,1}\frac{[H_{-n_{\Delta h},-n_C},H_{n_{\Delta h},n_C}]}{2n_{\Delta h}\Delta h+2n_C{\omega_c}}+O\left(\frac{\tilde{J_r}^3}{\Delta h^2}\right)+O\left(\frac{\tilde{J_r}^3}{{\omega_c}^2}\right)+O\left(\frac{\tilde{J_r}^3}{({\omega_c}-\Delta h)^2}\right)+O\left(\frac{\tilde{J_r}^3}{({\omega_c}+\Delta h)^2}\right)\\
&=&H_0+(\displaystyle{2\tilde{J}_r})^2\left[\displaystyle\frac{(a+a^\dagger)^2}{2\Delta h}-\frac{\Delta h}{{\omega_c}^2-\Delta h^2}(a^\dagger a+\frac{1}{2})\right]\sigma_z\\
\end{array}
\end{equation}
\end{widetext}
in a frame defined as $D=\exp{\left(-iS(t)\right)}$, with
\begin{equation}
\label{eq:VVframe}
S(t)=\displaystyle\sum_{n_{\Delta h},n_C}\frac{iH_{n_{\Delta h},n_C}}{2n_{\Delta h} \Delta h +2n_C{\omega_c}}F_{n_{\Delta h}}F_{n_C},
\end{equation}
 where $F_{n_{\Delta h}}=\exp{(2in\Delta h t)}$, $F_{n_C}=F_n=\exp{(2in{\omega_c} t)}$.  The appearance of $\sigma_z$ at this order suggests is the first non-QND term that arises (recalling that our scheme is performing measurements in the $\sigma_x$ basis, such that evolution due to $\sigma_z$ terms will causes deviations from the desired QND behavior).

We note that the Floquet resonances defined by the intuitive condition
\begin{equation}
\label{eq:resonances}
n_1 \Delta h + n_2 {\omega_c} + n_3 ({\omega_c}+\Delta h) + n_4 ({\omega_c} -\Delta h)\ll J_r/4
\end{equation}
with $|n_i-n_j|=\pm 1,0$ for $i,j=1,2,3,4$, are due to a skewed description of the system as a more appropriate description can be found in terms of slow envelopes of the remaining high-frequencies pulses. As a consequence, the usual high-frequency approximations in Floquet space can be supported by adiabatic considerations \cite{PhysRevA.95.023615,DeGrandi2010,Weinberg2017} leading to Eq. \ref{eq:VV}.

For completeness, it is also worth taking into consideration the tilting of the frame described in Eq. (\ref{eq:VVframe}) in which the Van Vleck Hamiltonian is valid. For example, at $t=0$, the change of frame is already non-trivial (although highly suppressed in the high frequency regime) and reads
\beq
S(0)=\displaystyle\sum_{n_{\Delta h},n_C}\frac{iH_{n_{\Delta h},n_C}}{2n_{\Delta h} \Delta h +2n_C{\omega_c}}.
\eeq By un-doing this change of frame with the operator $D_0=\exp{\left(-iS_0\right)}$ we get
\beq
H_0^{VV}&=&D^{-1}_0DHD^{-1}D_0,\\
&=&D^{-1}_0 H^{VV}D_0,\nonumber\\
&=&H^{VV}-i[H_0,S(0)]+O\left(\frac{g^3}{\Delta h^2}\right),\nonumber\\
&=&H^{VV}+\sum_{n_{\Delta h},n_C}\frac{[H_0,H_{n_{\Delta h},n_C}]}{2n_{\Delta h}\Delta h+2n_C{\omega_c}},\nonumber
\eeq
and, finally
\begin{widetext}
\begin{equation}
H^{VV}_0=H_0+({2\tilde{J}_r})^2\left[\frac{(a+a^\dagger)^2}{2\Delta h}-\frac{\Delta h}{{\omega_c}^2-\Delta h^2}(a^\dagger a+\frac{1}{2})-\frac{(a+a^\dagger)}{\Delta h}+\frac{\Delta h}{{\omega_c}^2-\Delta h^2}(1+(a+a^\dagger)^2)\right]\sigma_z
\end{equation}
\end{widetext}

The results of this analysis are collected in the following table.
\begin{widetext}
\renewcommand*{\arraystretch}{2.5}
\begin{center}
\resizebox{\columnwidth}{!}{%
\begin{tabular}{|c|c|c|c|c|}
\hline
\textbf{Range}&\textbf{Regime}&$H^{\text{RW}}$, $H^{VV}$&$H_{\text{eff}}$&\textbf{Error} \\
\hline
$\Delta h=0$&High-Freq.&$H_0^{\text{RW}}=2H_0$&$2H_0$&$O(\tilde{J_r}^2/{\omega_c})$\\
\hline
$0<\Delta h<\tilde{J_r}$&Adiabatic&$H_{\Delta h}^\text{RW}=H_0+{2\tilde{J}_r}(a+a^\dagger)(e^{2i\Delta h t}\sigma_++e^{-2i\Delta h t}\sigma_-)$&$2H_0$&$O(\Delta h)$\\
\hline
$\tilde{J_r}<\Delta h<\frac{{\omega_c}}{2}$&High-Freq.&$\begin{array}{cll}H_{\Delta h}^{\text{RW}}&=&H_0+{2\tilde{J}_r}(a+a^\dagger)(e^{2i\Delta h t}\sigma_++e^{-2i\Delta h t}\sigma_-)\\\ H^{\text{VV}}&=& H_0+({2\tilde{J}_r})^2\left[\frac{(a+a^\dagger)^2}{2\Delta h}-\frac{\Delta h}{{\omega_c}^2-\Delta h^2}(a^\dagger a+\frac{1}{2})\right]\sigma_z\end{array}$&$H_0$&$O(\tilde{J_r}^2/\Delta h)$\\
\hline
$\frac{{\omega_c}}{2}<\Delta h<{\omega_c}-\tilde{J_r}$&High-Freq.&$\begin{array}{cll}H_{\omega_c-\Delta h}^{\text{RW}}&=&H_0+{2\tilde{J}_r}[e^{2i(\omega_c-\Delta h)t}\sigma_- a^\dagger +e^{-2i(\omega_c-\Delta h)t}\sigma_+ a]\\ H^{\text{VV}}&=&H_0+({2\tilde{J}_r})^2\left[\frac{(a+a^\dagger)^2}{2\Delta h}-\frac{\Delta h}{{\omega_c}^2-\Delta h^2}(a^\dagger a+\frac{1}{2})\right]\sigma_z\end{array}$&$H_0$&$O\left(\tilde{J_r}^2/({\omega_c}-\Delta h)\right)$\\
\hline
${\omega_c}-\tilde{J_r}<\Delta h<{\omega_c}$&Adiabatic&$H_{{\omega_c}-\Delta h}^\text{RW}=H_{0}+{2\tilde{J}_r}[e^{2i(\omega_c-\Delta h)t}\sigma_- a^\dagger +e^{-2i(\omega_c-\Delta h)t}\sigma_+ a]$&$H_0$&$O({\omega_c}-\Delta h)+O(\tilde{J}_r)$\\
\hline
$\Delta h={\omega_c}$&High-Freq.&$H_{{\omega_c}}^\text{RW}=H_0+2\tilde{J_r}(\sigma_- a^\dagger+\sigma_+ a)$&$H_0$&$O(\tilde{J_r}^2/{\omega_c})+O(\tilde{J}_r)$\\
\hline
\end{tabular}}
\end{center}
\end{widetext}
In the table, $H^{VV}$ and $H^{\text{RW}}$ are the first order and rotating wave approximation to the full Hamiltonian, valid in the particular regime highlighted by the appropriate row in the table.  The Error column refers to the error occurring when one assumes $H\rightarrow H_{\text{eff}}$.

\bibstyle{apsrev4-1}
\bibliography{libraryLongReadout}

\end{document}